\documentclass[preprint2,numberedappendix]{emulateapj-rtx4}
\shorttitle{Current and magnetic helicity of active regions}
\shortauthors{Zhang et al.}
\usepackage{graphicx}
\usepackage{epsf}
\usepackage{epstopdf}
\newcommand{\bec}[1]{\mbox{\boldmath $ #1$}}
\begin{document}
\title{Current helicity of active regions as a tracer of large-scale solar magnetic helicity}

\author{H. Zhang}
\affil{National Astronomical Observatories, Chinese Academy of
Sciences, Beijing 100012, China}

\author{D. Moss}
\affil{School of Mathematics, University of Manchester,
Manchester M13 9PL, UK}

\author{N. Kleeorin}
\affil{Department of Mechanical Engineering, Ben-Gurion University
of Negev, POB 653, 84105 Beer-Sheva, Israel}
\affil{NORDITA, AlbaNova University Center,
Roslagstullsbacken 23, SE-10691 Stockholm, Sweden}

\author{K. Kuzanyan}
\affil{IZMIRAN, Troitsk, Moscow Region 142190, Russia}

\author{I. Rogachevskii}
\affil{Department of Mechanical Engineering, Ben-Gurion University
of Negev, POB 653, 84105 Beer-Sheva, Israel}
\affil{NORDITA, AlbaNova University Center,
Roslagstullsbacken 23, SE-10691 Stockholm, Sweden}

\author{D. Sokoloff}
\affil{Department of Physics, Moscow State University, Moscow
119992, Russia}

\and

\author{Y. Gao and H. Xu}
\affil{National Astronomical Observatories, Chinese Academy of
Sciences, Beijing 100012, China}





\begin{abstract}
We demonstrate that the current helicity observed
in solar active regions traces the magnetic
helicity of the large-scale dynamo generated
field. We use an advanced 2D mean-field dynamo
model with dynamo saturation based on the
evolution of the magnetic helicity and algebraic
quenching. For comparison, we also studied a more
basic 2D mean-field dynamo model with simple
algebraic alpha quenching only. Using these
numerical models we obtained butterfly diagrams
both for the small-scale current helicity and
also for the large-scale magnetic helicity, and
compared them with the butterfly diagram for the
current helicity in active regions obtained from
observations. This comparison shows that the
current helicity of active regions, as estimated
by $-{\bf A \cdot B}$ evaluated at the depth from
which the active region arises, resembles the
observational data much better than the
small-scale current helicity calculated directly
from the helicity evolution equation. Here ${\bf
B}$ and ${\bf A}$
are respectively the dynamo
generated mean magnetic field and its vector
potential. A theoretical interpretation of these
results is given.
\end{abstract}


\section{Introduction}

The solar activity cycle is believed to be a
manifestation of dynamo action which somewhere in
the solar interior generates waves of
quasi-stationary magnetic field propagating from
middle latitude towards the solar equator
("dynamo waves"). The traditional explanation of
this dynamo action \citep{par55} is based on  the
joint action of differential rotation and mirror
asymmetric convection which results in what has
come to be known as the $\alpha$-effect, based on
the helicity of the hydrodynamic convective flow
\citep{KR80,M78}. This explanation is however not
the only one discussed in the literature and, for
example,  meridional circulation is also
suggested as an important co-factor of the
$\alpha$-effect, see e.g., \cite{dg01,cetal04}.

In turn, traditional dynamo scenarios based on
differential rotation and the classical
$\alpha$-effect have to include a dynamo
saturation mechanism. One of the most popular
saturation mechanisms is based on a contribution
to the $\alpha$-effect from magnetic fluctuations
\citep{petal76}. A relevant quantification of
this effect involves considerations of magnetic
helicity evolution, e.g., \cite{KRR95,ketal03}. A
key role in the evolution of magnetic helicity is
played by magnetic helicity fluxes
\citep{ketal00,bf00,BS05}. Of course, this
scenario is not the only one that has been
suggested: for example \cite{bf00b,bb03,b07}
consider coronal-mass ejections as an important
part of nonlinear suppression of the dynamo, and
\cite{metal11a} discuss the effects of losses via
the solar wind. \cite{cetal04} believe that the
current helicity in solar active regions is
substantially modified when magnetic tubes rise
up to the solar surface.

A natural way to resolve such controversies is to determine relevant
quantities such as the $\alpha$-effect through observations,
this providing a check on the various scenarios.
Such an option is now becoming
realistic, starting from the 1990s when  the first attempts to
observe current helicity in solar active regions were undertaken
\citep{s90,petal94,bz98,hs04}.

Twenty years of continuous efforts by several observational groups,
with the main contribution coming from the Huairou Solar Station of
China, has resulted \citep{zetal10} in reconstruction of the current
helicity time-latitude (butterfly) diagram for one full solar
magnetic cycle (1988-2005). From this butterfly diagram it is
apparent that the current helicity is involved in the solar activity
cycle and follows a polarity law comparable with the Hale polarity
law for sunspots -- but rather more complicated. In other words,
dynamo generated magnetic field is indeed mirror asymmetric and this
mirror asymmetry is involved in the solar activity cycle and can be
used to understand its nature \citep{ketal03,zetal06}.

What however needs clarification is which dynamo governing parameter
is traced by such a surface proxy as a measure of current helicity
in solar active regions. A naive idea here is to identify this part
of the surface current helicity with the current and magnetic
helicities of the dynamo generated small-scale magnetic fields deep
inside the Sun (say, in the solar overshoot layer), which suppress
dynamo action. To start with something definite, this naive
interpretation was applied by \cite{ketal03} and \cite{zetal06}.

Recent progress in observation which has resulted in butterfly
diagrams for the current helicity in active regions \citep{zetal10}
makes it possible to go further with this topic. The aim of this
paper is to argue that the current helicity in solar active regions
directly reflects {\it the magnetic
helicity of the large-scale dynamo generated field.} We organize our
arguments as follows.

Based on the observed butterfly diagram for the current helicity, we here
confront the naive interpretation with the available observations. We examine
a mean-field dynamo model
with dynamo suppression based on the magnetic helicity balance,
obtain the corresponding current helicity butterfly diagrams,
and also that of the large-scale magnetic helicity, and compare
them with those observed. This comparison shows that
the evolution of the large-scale magnetic helicity resembles
the observational data much more closely than that of the current helicity
of the small-scale fields.
In order to demonstrate the robustness of this result,
we also consider a more primitive dynamo model, with a simple algebraic
$\alpha$-quenching. From this model we calculate the large-scale magnetic
helicity and compare it with
the observational butterfly diagram.
We find that this fits observations more or less as well  as that from
a dynamo model based on helicity conservation.
We conclude that the major part of the observed current
helicity in active regions is produced in the rise of magnetic
loops to the solar surface.

\section{The role of helicities in magnetic field evolution}

In the evolution of the magnetic field different
helicities play different roles. Considering the
small-scale velocity and magnetic fluctuations,
${\bf u}$ and ${\bf b}$ respectively, there are
three helicities: (i) the kinetic helicity $H_u =
\langle {\bf u} {\bf \cdot} \bec{\rm curl} \,{\bf
u} \rangle$ that determines the kinetic $\alpha$
effect; (ii) the current helicity $H_c=\langle
{\bf b} {\bf \cdot} \bec{\rm curl} \,{\bf b}
\rangle$ that determines the magnetic part of the
$\alpha$ effect; and (iii) the magnetic helicity
$H_m =\langle {\bf a} {\bf \cdot} {\bf b}
\rangle$, where ${\bf b}=\bec{\rm curl} \,{\bf
a}$. We use the angled brackets $\langle \ldots
\rangle$ to  denote spatial integrals over all
relevant turbulent fluid. These integrated
helicities are used in the sense of average
helicity densities.

The total magnetic helicity, the sum $H_M+H_m$
of the magnetic helicities of the large- and
small-scale fields, $H_M={\bf A} {\bf \cdot} {\bf
B}$ and $H_m$ respectively, is conserved for very
large magnetic Reynolds numbers. Here ${\bf
B}=\bec{\rm curl} \,{\bf A}$ is the large-scale
magnetic field. Note that, on the contrary, the
current helicity $H_c$ is not conserved. On the
other hand, the kinetic helicity $H_u$ is
conserved only for very large fluid Reynolds
numbers when the large-scale magnetic field ${\bf
B}$ vanishes. The characteristic time for the
decay of kinetic helicity is of the order of the
turn-over time $\tau=\ell / u$ of turbulent
eddies in the energy-containing scale, $\ell$, of
turbulence, while the characteristic time of the
small magnetic helicity decay is of the order of
$T_m =\tau {\rm Rm}$ \citep{M78,ZRS83,BS05},
where ${\rm Rm} = \ell \, u /\eta_{_{0}}$ is the
magnetic Reynolds number, $u$ is the
characteristic turbulent velocity and
$\eta_{_{0}}$ is the magnetic diffusivity due to
electrical conductivity of the fluid. The
small-scale current helicity, $H_c$ is not an
integral of motion and the characteristic time of
$H_c$ varies from a short timescale, $\tau$, to
much larger timescales. On the other hand, the
characteristic decay times of the current
helicity of large-scale field, $H_C={\bf B} {\bf
\cdot} \bec{\rm curl} \,{\bf B}$, and of the
large-scale magnetic helicity, $H_M$, are of the
order of the turbulent diffusion time. For weakly
inhomogeneous turbulence the small-scale current
helicity, $H_c$, is proportional to the
small-scale magnetic helicity, $H_m$.

As the dynamo amplifies the large-scale magnetic
field, the large-scale magnetic helicity
$H_M={\bf A} {\bf \cdot} {\bf B}$ grows in time
(but not monotonically in a cyclic system). The
evolution of the large-scale magnetic helicity
$H_M$ is determined by
\begin{eqnarray}
{\partial H_M \over \partial t} + \bec{\nabla} \cdot {\bf F}_M
= 2 \bec{\cal E} \cdot {\bf B} - 2 \eta H_C \;,
\label{H1}
\end{eqnarray}
\citep{KRR95,bf00,BS05}, where $\bec{\cal E} =
\langle {\bf u}{\bf \times}{\bf b} \rangle$  is
the mean electromotive force that determines
generation and dissipation of the large-scale
magnetic field, $2 \bec{\cal E} \cdot {\bf B}$ is
the source of the large-scale magnetic helicity
due to the dynamo generated large-scale magnetic
field, and ${\bf F}_M$ is the flux of large-scale
magnetic helicity that determines its transport.
Since the total magnetic helicity over all
scales, $H_M + H_m$ integrated over the turbulent
fluid, is conserved for very small magnetic
diffusivity, the small-scale magnetic helicity
changes during the dynamo process, and its
evolution is determined by the dynamic equation
\begin{eqnarray}
{\partial H_m \over \partial t} + \bec{\nabla} \cdot {\bf F}
= - 2 \bec{\cal E} \cdot {\bf B} - 2 \eta H_c \;,
\label{H2}
\end{eqnarray}
\citep{KRR95,bf00,BS05}, where $- 2 \bec{\cal E}
\cdot {\bf B}$ is the source of the small-scale
magnetic helicity due to the dynamo generated
large-scale magnetic field,  ${\bf F}$ is the
flux of small-scale magnetic helicity that
determines its transport and $2 \eta H_c = H_m /
T_m$ is the dissipation rate of the small-scale
magnetic helicity. It follows from Eq.~(\ref{H1})
and~(\ref{H2}) that the source of the small-scale
and the large-scale magnetic helicities is
located only in turbulent regions (i.e., in our
case, in the solar convective zone). The magnetic
part of the $\alpha$ effect is determined by the
parameter $\chi^{c}= \tau H_c/(12 \pi \rho)$, and
for weakly inhomogeneous turbulence $\chi^{c}$ is
proportional to the magnetic helicity: $\chi^{c}
= H_{m} / (18 \pi \eta_{_{T}} \rho)$
\citep{KR99,BS05},  where $\rho$ is the density
and $\eta_{_{T}}$ is the turbulent magnetic
diffusion.

\section{The observed current helicity butterfly diagrams}

\begin{figure*}
\begin{center}
\includegraphics[height=0.86\textwidth,angle=90]{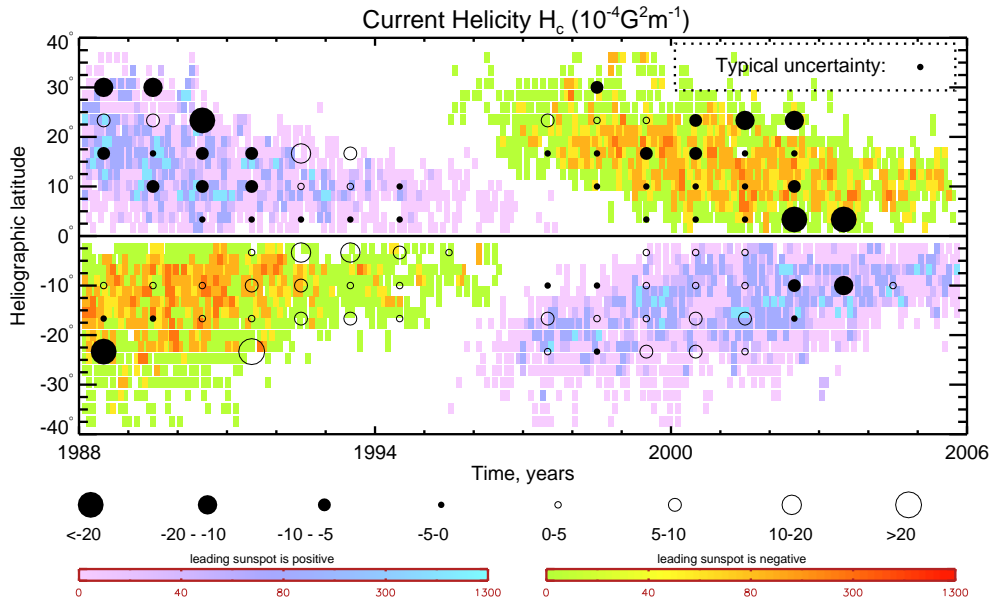}
\caption{Observed current helicity (white/black
circles for positive/negative values) for solar
active regions in the 22$^{nd}$ and 23$^{rd}$
solar cycles as averaged over two-year running
windows over latitudinal bins of 7 degrees wide,
overlaid with sunspot density (colour). The
circle in the upper right corner of the panel
indicates the typical value of observational
uncertainty defined by 95\% confidence intervals
scaled to the same units as the circles. The
vertical axis gives the latitude in degrees and
the horizontal gives the time in years. }
\label{Fig0}
\end{center}
\end{figure*}

The observed butterfly diagrams of electric
current helicity for solar active regions during
the last two solar cycles has been presented by
\cite{zetal10}. The general structure can be
described as follows. Current helicity is
involved in the solar activity cycle and follows
a polarity rule comparable  to (however more
complicated than) the polarity rule for toroidal
magnetic field which in turn comes from the Hale
polarity rule for sunspot groups. Migration of
the helicity pattern is clearly visible and
located near the toroidal field pattern. The
wings of the helicity butterflies are slightly
more inclined to the equator than the magnetic
field wings, but the former follow in general the
latter. As a quantity quadratic in magnetic
field, the current helicity in one hemisphere has
the same sign for both cycles, with the opposite
sign in the other hemisphere (a kind of
unchanging dipolar symmetry).

Fig.~1 shows the distribution of the average helical characteristics
of the magnetic field in solar active regions in the form of
a butterfly diagram (latitude-time) for 1988-2005 (which covers
most of the 22nd and 23rd solar cycles). These results are
inferred from photospheric vector magnetograms recorded at Huairou
Solar Observing Station.

This, the longest available systematic dataset
covering the period of two solar cycles,
comprises 6205 vector magnetograms of 984 solar
active regions (most of the large solar active
regions of both solar cycles). Of these, 431
active regions belong to the 22nd solar cycle and
553 to the 23rd. We have limited the latitudes of
active regions to $\pm$40$^{\circ}$ and most of
them are below $\pm$35$^{\circ}$. The helicity
values of the active regions have been averaged
over latitude by intervals of 7$^{\circ}$ in
solar latitude, and over overlapping two-year
periods of time (i.e., 1988-1989, 1989-1990, ...,
2004-2005). By this method of averaging we were
able to group sets of at least 30 data points in
order to make error estimations (computed as 95
\% confidence intervals) reasonably small. In
this sampling we find that 66\% (63\%) of active
regions have negative (positive) mean current
helicity in the northern (southern) hemisphere
over the 22nd solar cycle and 58$\%$ (57\%) in
the 23rd solar cycle.

There are some domains in the diagram where
current helicity has  the "wrong" sign with
respect to the global polarity law. These domains
of "wrong" sign are located at the very beginning
and the very end of the wings.

Concerning alternative explanations of the
observations \citep{zetal10}, \cite{MB05,YM09}
describe current helicity in solar active regions
in terms of generation of non-potential coronal
structures by surface differential rotation. Note
that the surface differential rotation cannot
generate large-scale and small-scale magnetic
helicity. It can only redistribute the existing
magnetic helicity, as can any non-uniform
large-scale motions. This process is determined
by the flux term in the evolutionary equation for
the magnetic helicity. The local change of the
magnetic helicity inside a given active region by
the surface differential rotation plays an
important role. In our paper we study robust
global (rather than local) features of the
evolution of the magnetic helicity, by averaging
over an ensemble of active regions. In this case
the global evolution of the magnetic helicity
mainly depends on the sources of magnetic
helicity inside the solar convective zone.

\section{An estimate for the current helicity in active regions}

In this section we estimate the current helicity in
active regions. There is a common belief that active regions are
formed due to some nonaxisymmetric instability
of $\sim 100$ kG magnetic fields in the tachocline, e.g., \citep{GD00,CDG03,PM07}.
However, the existence of such strong fields and the role of this mechanism
 remain questionable \citep{B05}.
Another promising mechanism of formation of
active regions  is related to a negative
effective magnetic pressure instability of the
large-scale dynamo generated magnetic field. This
instability was predicted theoretically
\citep{KMR96,RK07} and detected recently in
direct numerical simulations \citep{BKKMR11}. The
instability is caused by the suppression of
turbulent hydromagnetic pressure by the mean
magnetic field. At large Reynolds numbers and for
sub-equipartition magnetic fields, the resulting
negative turbulent contribution can become so
large that the effective mean magnetic pressure
(the sum of turbulent and non-turbulent
contributions) appears negative
\citep{BKR10,BKKMR11}. In a stratified turbulent
convection, this results in the excitation of a
large-scale instability that results in the
formation of large-scale inhomogeneous magnetic
structures. This mechanism is consistent with the
suggestion that active regions are formed near
the surface of convective zone \citep{k10}.

The spatial scale of an active region is much smaller than
the solar radius, but  much larger then the maximum scale
of solar turbulent granulation. To estimate the current
helicity in an active region, we
have to relate the large-scale magnetic field $\bf B$
and its magnetic potential $\bf A$ inside the convective
zone as well as the corresponding small-scale quantities
inside the convective zone (which
determine the small-scale magnetic fluctuations),
with the surface magnetic field ${\bf B}^{\rm ar}$ and its magnetic
potential ${\bf A}^{\rm ar}$ inside active regions, which are
the quantities available to observations.

We base our estimate on the following idea.
Consider a new-born magnetic tube at the layer in
convective zone from which the active region is
developing. We assume that the rise of magnetic
tubes is a fast adiabatic process. Let us also
assume that the mean magnetic field and the total
magnetic helicity vanish at the initial instant,
and take into account the magnetic helicity
conservation law (as solar plasma is highly
conductive, and so we consider the magnetic
helicity conservation law to hold at all scales,
including that of the whole Sun). If this tube
rises rapidly to the surface to produce an active
region, the total magnetic helicity in the tube
is conserved because the process is rapid. Rising
large-scale magnetic field and magnetic potential
give the corresponding quantities for active
regions, which may thus differ substantially from
the corresponding quantities in the surrounding
medium. Because the initial total magnetic
helicity of the tube, which was almost
nonmagnetized, was negligible, the magnetic
helicity conservation law reads
\begin{eqnarray}
\langle {\bf A}^{\rm ar}{\bf\cdot}{\bf B}^{\rm ar}
\rangle \approx - {\bec{ A}} {\bf \cdot} {\bec{ B}},
\label{R5}
\end{eqnarray}
where the angular brackets denote averaging over the
surface occupied by the active region.

Now we relate the mean current helicity $\langle
{\bf B}^{\rm ar} {\bf \cdot} \bec{\rm curl}
\,{\bf B}^{\rm ar} \rangle$  with the magnetic
helicity $\langle {\bf A}^{\rm ar}{\bf \cdot}{\bf
B}^{\rm ar}\rangle$. We rewrite it from the first
principles with the use of permutation tensors as
\begin{eqnarray}
\langle {\bf B}^{\rm ar} {\bf \cdot} \bec{\rm curl} \,{\bf B}^{\rm ar}
\rangle \approx {1\over L^2_{\rm ar}} \, \langle {\bf A}^{\rm ar}{\bf \cdot}{\bf B}^{\rm ar}
\rangle + O\biggl({L^2_{\rm ar} \over R_\odot^2}\biggr),
\label{RR5}
\end{eqnarray}
(see Appendix \ref{CH-MH}), where $R_\odot$ is the solar radius, and
$L_{\rm ar}$ is the spatial scale  of an active region.
Equations~(\ref{R5}) and~(\ref{RR5}) yield
\begin{eqnarray}
\langle {\bf B}^{\rm ar} {\bf \cdot} \bec{\rm curl} \,{\bf B}^{\rm ar} \rangle
\approx - {1 \over L^2_{\rm ar}} {\bec{ A}} {\bf
\cdot} {\bec{ B}} .
\label{RRR5}
\end{eqnarray}
Therefore, the observed current helicity in
active regions is expected to be a proxy for
$-{\bf A} \cdot {\bf B}$. This idea will be
checked using mean-field dynamo numerical
modelling and comparison of the numerical results
with the observed current helicity in active
regions.

\section{Dynamo models}

Our approach to compare the dynamo models with observations is as
follows. We consider two types of dynamo models. Both types of models
are 2D mean-field models with an axisymmetric magnetic field which
depends on radius $r$ and polar angle $\theta$.
The third (azimuthal) coordinate is $\phi$, and $\partial/\partial\phi=0$.
The dynamo action is based on differential rotation, with a
rotation curve which resembles that of the solar convection zone,
as known from
helioseismological observations, and there is a conventional $\alpha$-effect.

The first type of model assumes a very naive
algebraic $\alpha$-quenching.  Then we suppose
that the total magnetic helicity is locally
vanishing, so the magnetic helicity of the
large-scale magnetic field produced in the
course of mean-field dynamo action has to be
compensated by small-scale magnetic helicity.
(Thus we are assuming that at an initial instant
the medium is non-magnetic, so that helicity
conservation means  that the sum of large and
small-scale helicities remains zero.) We assume
also that there is a separation of scales so that
characteristic turbulence scales are much smaller
than the characteristic spatial scales of mean
magnetic field variations. This allows a link
between current and magnetic helicities
\citep{KR99} to be made. This concept underlies
the observational procedure for determining the
current helicity of active regions, and for
calculating the current helicity from the
magnetic helicity of the small-scale fields.
Based on the same concept we estimate the
large-scale magnetic helicity as $B_\phi A_\phi$,
where $A(r,\theta)\hat\phi$ is the magnetic
potential for the poloidal field. As a result we
obtain (for a given radius $r$) a theoretical
model for the current helicity as a function of
$t$ and $\theta$ which we  overlay on the
butterfly diagram for $B_\phi$. We compare the
result with the current helicity butterfly
diagram known from observations and obtained
using  similar underlying concepts.

A further point is that the primitive model
allows a simplification to the level of the
one-dimensional Parker migratory dynamo, and this
opportunity has been investigated in this respect
by \cite{xetal09}. We will use the results of
that work for reference and comparison.

We do not consider this primitive scheme as
realistic. We are sure that any more or less
realistic scenario for solar dynamo suppression
will have to be much more sophisticated. On the
other hand, we can see whether this primitive
model produces a helicity butterfly diagram that
is quite similar to that observed. The only
shortcoming of the model is that the maximal
current helicity occurs later then the maximum
$B_\phi$, while it is observed to come up
earlier. If magnetic helicity conservation
determines the nonlinear dynamo suppression, we
expect that a careful reproduction of this
balance, including helicity fluxes and the link
between magnetic helicity and $\alpha$-effect,
will result in even a better theoretical
butterfly diagram, and possibly improve the phase
relations between helicity and toroidal magnetic
field.

As a specific example of the second type model
that takes into account the influence of magnetic
helicity balance on dynamo action we use the
dynamo model described by \cite{zetal06}. Whereas
simple $\alpha$-quenching provides a quite robust
suppression of a spherical dynamo and give (more
or less) steady nonlinear magnetic field
oscillations for a very wide range of parameters,
in contrast it is far from clear {\it a priori}
that a dynamo suppression based on magnetic
helicity conservation is effective enough to
suppress magnetic field growth and result in
steady oscillations. In fact it works
more-or-less satisfactorily only in a quite
narrow parameter range \citep{ketal03,zetal06},
which appears inadequate to fit observations.

We note two crucial points here. First of all, both types of  models
ignore any direct action of magnetic force on the rotation law.
In the more primitive models, there is a crude parametrization
of feedback onto the (purely hydrodynamic) alpha effect [see Eq.~(\ref{helic-prim})]
below. The latter more sophisticated model describes the back-reaction of the
generated magnetic field on the dynamo process in terms
of the magnetic contribution of the current
helicity onto the magnetic part of $\alpha$-effect. On the other hand,
the feedback of the generated large-scale magnetic field on
turbulent convection is described in our model by the algebraic
quenching of $\alpha$-effect, pumping velocities and turbulent
magnetic diffusion.

We assume that helicity conservation is
not the only mechanism of dynamo suppression. The fact that we
see a manifestation of helicity on the solar surface tells us that the
buoyancy indeed plays some role, and we add it to the model. We
stress that the buoyancy which we include in the model transports
current helicity and magnetic helicity as well as large-scale
magnetic field.

Below we discuss the detailed dynamo models.
We use spherical coordinates $r, \theta, \phi$
and describe an axisymmetric magnetic field by the azimuthal
component of magnetic field $B$, and the component $A$ of
the magnetic potential corresponding to the
poloidal field.

We measure length in units of the solar radius $R_\odot$, and
time in units of a diffusion time based on the solar
radius and the reference turbulent magnetic diffusivity $\eta_{_{T0}}$.
The magnetic field is measured in units of the equipartition field
$ B_{\rm eq} = u_\ast \,  \sqrt{4 \pi \rho_\ast}$,  the vector
potential of the poloidal field $A$ in units of
$R_\odot B_{\rm eq}$, the density $\rho$ is normalized with
respect to its value $\rho_\ast$ at the bottom of the convective
zone, and the basic scales of the turbulent motions $\ell$ and
turbulent velocity $u$ at the scale $\ell$ are measured in units of
their maximum values through the convective zone. The $\alpha$-effect
is measured in terms of $\alpha_0$, defined below, and angular velocity
in units of the maximum surface value, $\Omega_0$.

\subsection{The primitive, alpha-quenched model}
\label{primitive}

In the primitive dynamo model the $\alpha$-effect is given by
\begin{equation}
\alpha = \alpha^v = \chi^v \Phi_v \;,
\label{helic-prim}
\end{equation}
where $\chi^v$ is proportional to the
hydrodynamic helicity, $H_u$, multiplied by the
turbulent correlation time $\tau$, and $\Phi_v
=(1 + B^2)^{-1}$ is the model for the $\alpha$
quenching nonlinearity. For convenience, we use
for most of these computations the code of
\cite{mb00}  -- see also \cite{metal11b}. This
code has the possibility of a modest reduction in
the diffusivity, to $\eta_{\rm min}$, in the
innermost part of the computational shell
("tachocline"), below fractional radius 0.7. We
define $\eta_{\rm r}=\eta_{\rm
min}/\eta_{_{T0}}$. We also used this primitive
formulation of alpha-quenching in the (otherwise
very similar) model used in Sect.~\ref{helbal}
when producing Fig.~\ref{Fig4}. In the latter
case, the diffusivity is everywhere uniform.

At the surface $r=1$ the field is matched to a vacuum external field,
and "overshoot" boundary conditions are used at the lower boundary.

\subsection{The model based on helicity balance}
\label{helbal}

Here we use the code described in \cite{zetal06}, with two new
features: we allow the possibility of meridional circulation and/or
vertical motions attributed to magnetic buoyancy.
The dynamo equations for $\tilde A = r \sin \theta A$ and
$\tilde B = r \sin \theta \, B$ read
\begin{eqnarray}
&& {\partial \tilde  A \over \partial t} + {(V^A_\theta + V^M_\theta) \over r}
   {{\partial \tilde A} \over {\partial \theta}} + V_r \,
   {{\partial \tilde A} \over {\partial r}} = C_\alpha \, \alpha \,
   \tilde B
   \nonumber \\
&& \; \; \; \; + \eta_{_{A}} \biggl[{{\partial ^2 \tilde A }
   \over {\partial r^2}} + {\sin \theta \over r^2} {\partial
   \over \partial \theta} \biggl({1 \over \sin \theta} {\partial
   \tilde A \over \partial \theta}\biggr) \biggr] \;,
\label{L1} \\
&& {{\partial \tilde B} \over {\partial t}} +  {\sin \theta \over
   r} {\partial \over \partial \theta} \biggl[{(V^B_\theta + V^M_\theta) \tilde B
   \over \sin \theta}\biggr] + {{\partial [V_r \, \tilde B]} \over
   {\partial r}}
   \nonumber \\
&&  \; \; \; \;  = C_\omega \,  \sin \theta \, \biggl[ G_r {{\partial } \over
   {\partial \theta}} - G_\theta {{\partial } \over {\partial r}}
   \biggr] \tilde A + {\sin \theta \over r^2} {\partial \over
   \partial \theta} \biggl[{\eta_B \over \sin \theta} {\partial
   \tilde B \over \partial \theta}\biggr]
   \nonumber \\
&&  \; \; \; \;
   + {\partial \over {\partial r}} \biggl[ \eta_{_{B}}
{{\partial \tilde B} \over {\partial r}} \biggr] \;,
\label{L2}
\end{eqnarray}
where $V_r=V^A_r + V^M_r+V_B$,  $\,
\bec{V}^{A}(B)$ and $\bec{V}^{B}(B)$ are the
nonlinear drift velocities of poloidal and
toroidal mean magnetic fields, $\bec{V}^{M}$ is
the meridional circulation velocity, ${\bf V_{\rm
B}}$ is the vertical buoyancy velocity,
$\eta_{_{A}}(B)$ and $\eta_{_{B}}(B)$ are the
nonlinear turbulent magnetic diffusion
coefficients for the mean poloidal and toroidal
magnetic fields,  and and the non-dimensional
dynamo parameters are $C_\alpha=\alpha_0 \,
R_\odot / \eta_{_{T0}}$, $\, C_\omega=\Omega_0 \,
R_\odot^2 / \eta_{_{T0}}$. The nonlinear
turbulent magnetic diffusion coefficients and the
nonlinear drift velocities are given in Appendix
\ref{Dyn-mod}. The non-dimensional gradients of
differential rotation are

\begin{eqnarray*}
G_r = {{\partial \Omega} \over {\partial r}} \;, \quad G_\theta =
{{\partial \Omega} \over {\partial \theta}} .
\end{eqnarray*}
In this dynamo model with magnetic helicity evolution
the total $\alpha$-effect is given by
\begin{equation}
\alpha = \alpha^v + \alpha^m = \chi^v \Phi_v
+ {{ \Phi_m} \over {\rho(z)}}\chi^c \;,
\label{helic}
\end{equation}
with $\alpha^v=\alpha_0\sin^2\theta\cos\theta \, \Phi_v$.
The magnetic part of the $\alpha$-effect is based
on the idea of magnetic helicity conservation and the link between
current and magnetic helicities. Here $\chi^v$ and $\chi^c$
are proportional to the hydrodynamic
and current helicities multiplied by the turbulent correlation
time, and $\Phi_v$ and $\Phi_m$ are
quenching functions. The analytical form of the quenching
functions $\Phi_{v}(B)$ and $\Phi_{m}(B)$ is given in Appendix \ref{Dyn-mod}.
The density profile is chosen in the form:
\begin{equation}
\rho(z) = \exp [- a \tan (0.45 \pi \, z)] \;, \label{rho}
\end{equation}
where $z=1 - \mu (1-r) $ and $\mu = (1 - R_0/R_\odot)^{-1}$. Here
$a \approx 0.3$ corresponds to a tenfold change of the density in
the solar convective zone, $a \approx 1$ by a factor of about $10^3$.

The equation for $\tilde \chi^c = r^2 \sin^2 \theta \, \chi^c$ is
\begin{eqnarray}
&& {{\partial \tilde \chi^c} \over {\partial t}} + {{\tilde\chi^c}
\over T} = \left({{2R_\odot} \over \ell}\right)^2 \biggl\{ {1 \over
C_\alpha} \biggl[{\eta_{_{B}} \over r^2} \, {{\partial \tilde A}
\over {\partial \theta}} {{\partial \tilde B} \over {\partial
\theta}} + \eta_{_{B}} \, {{\partial \tilde A} \over {\partial r}}
{{\partial \tilde B} \over {\partial r}}
\nonumber\\
&& \; \; - \eta_{_{A}} \, \tilde B \, {\sin \theta \over r^2}
{\partial \over \partial \theta} \biggl({1 \over \sin \theta}
{\partial \tilde A \over \partial \theta}\biggr) - \eta_{_{A}} \,
\tilde B {{\partial ^2 \tilde A} \over {\partial r^2}}
\nonumber\\
&& \; \; + (V^A_r - V^B_r) \, \tilde B {{\partial \tilde A} \over
{\partial r}} + (V^A_\theta - V^B_\theta) {\tilde B \over r}
{{\partial \tilde A} \over {\partial \theta}} \biggr] - \alpha
\tilde B^2 \biggr\}
\nonumber\\
&& \; \; - {{\partial [\tilde {\cal F}_r + (V_B + V^M_r) \,
\tilde \chi^c]} \over {\partial r}} -
{\sin \theta \over r} {{\partial} \over {\partial \theta}}
\left[{\tilde {\cal F}_\theta + V^M_\theta \, \tilde \chi^c
\over  \sin \theta} \right]  \;,
\nonumber\\
\label{curhel}
\end{eqnarray}
where $\bec{\tilde {\cal F}} = r^2 \sin^2 \theta
\bec{\cal F}$, $\, \bec{\cal F}= {\bf F}/ (18 \pi
\eta_{_{T}} \rho)$ is related to the flux ${\bf
F}$ of the small-scale magnetic helicity and
given in Appendix \ref{Dyn-mod}, $R_\odot/\ell$
is the ratio of the solar radius to the basic
scale of solar convection, $ T = (1/3) \, {\rm
Rm} \, (\ell/R_\odot)^2 $ is the dimensionless
relaxation time of the magnetic helicity, ${\rm
Rm} = \ell \, u / \eta_{_{0}}$ is the magnetic
Reynolds number, with $\eta_{_{0}}$ the `basic'
magnetic diffusion due to the electrical
conductivity of the fluid.

The meridional circulation (single cell in each hemisphere, poleward at surface)
is determined by
\begin{eqnarray}
V^M_\theta &=& - {1 \over \sin \theta \, r \, \rho(r)} {{\partial [r \, \Psi(r,\theta)]}
\over {\partial r}}\;,
\label{mer1}\\
V^M_r &=& {1 \over \sin \theta \, r \, \rho(r)} {{\partial
\Psi(r,\theta)} \over {\partial \theta}}\;,
\label{mer2}
\end{eqnarray}
where $\Psi(r,\theta) = R_v \,  \sin^2\theta\cos\theta f(r) \rho$,
$\, f(r)=2(r-r_b)^2(r-1)/(1-r_b)^2$, $\, r_b$ is the base of
the computational shell.
This is normalized so that the max of $V^M_\theta$ at the
surface is unity.
We introduce a coefficient $R_v=R_\odot \, U_0 / \eta_{_{T0}}$,
where $U_0$ the maximum surface speed.

Buoyancy is implemented by the introduction of a
purely vertical velocity ${\bf V_{\rm B}}=\gamma
B_\phi^2\hat{\bf r}$ in Eqs.~(\ref{L1}),
(\ref{L2}) and~(\ref{curhel}), where $\gamma>0$
\citep{metal99}. We justify the apparent
non-conservation of mass by adopting the argument
of K.-H.~R\"adler, presented as a private
communication in \cite{metal90}, that the return
velocity will be in the form of a more-or-less
uniform "rain". In some ways the process
represents pumping by a "fountain flow". As a
result the regular velocity ${\bf V}_B$ appears
in the governing equations for the large-scale
magnetic field and magnetic and current
helicities but not the the equation for density.
From the viewpoint of probability theory, in the
first case ${\bf V}_B$ is a mean quantity taken
under the condition that an elementary volume is
magnetized so it does not vanish, while in the
second case this mean is taken without any
condition and vanishes. In our opinion, this idea
can also be constructive for other problems with
magnetic helicity advective fluxes, e.g.
\cite{setal06}.

At the surface of the Sun, $r=1$, we use vacuum boundary conditions
on the field, i.e. $B=0$ and the poloidal field fits smoothly onto a
potential external field. At the lower boundary (the bottom of the
solar convective zone), $r=r_0=0.64$, $\, B=B_r=0$.
At both $r=r_0$ and $r=1$,  we set $\partial \chi^c/\partial
r=0$, where $\chi^c$ is proportional to the current helicity [see
Eq.~(\ref{helic})].

\section{Simulated butterfly diagrams for current helicity}

We performed an extensive numerical investigation of the models
in a parametric range which is considered to be adequate for solar
dynamos. We estimate the values of the governing parameters for
different depths of the convective zone, using models of the solar
convective zone, e.g. \cite{bt66}, \cite{s74} -- more modern
treatments make little difference to these estimates. In the upper
part of the convective zone, say at depth (measured from the top) $
h_\ast = 2 \times 10^7$ cm , the parameters are $ {\rm Rm} = 10^5 ,$
$\, u = 9.4 \times 10^4 $ cm s$^{-1}$, $ \ell = 2.6 \times 10^7$ cm,
$ \, \rho = 4.5 \times 10^{-7}$ g cm$^{-3} ,$ the turbulent
diffusivity $\, \eta_{_{T}} = 0.8 \times 10^{12} $ cm$^2$ s$^{-1}$;
the equipartition mean magnetic field is $B_{\rm eq} = 220 $ G and $
T = 5 \times 10^{-3}$. At depth $ h_\ast = 10^9$ cm these values are
$ {\rm Rm} = 3 \times 10^7$, $\, u = 10^4$ cm s$^{-1}$, $\, \ell =
2.8 \times 10^8$ cm, $\, \rho = 5 \times 10^{-4}$ g cm$^{-3} ,$ $\,
\eta_{_{T}} = 0.9 \times 10^{12} $ cm$^2$ s$^{-1}$; the
equipartition mean magnetic field is $B_{\rm eq} = 800$ G and $T
\sim 150$. At the bottom of the convective zone, say at depth
$h_\ast = 2 \times 10^{10}$ cm, $\, {\rm Rm} = 2 \times 10^9 ,$ $ \,
u = 2 \times 10^3 $ cm s$^{-1}$, $\ell = 8 \times 10^9$ cm, $ \rho =
2 \times 10^{-1}$ g cm$^{-3} ,$ $\, \eta_{_{T}} = 5.3 \times 10^{12}
$ cm$^2$s$^{-1}$. Here the equipartition means magnetic field
$B_{\rm eq} = 3000 $ G and $T \approx 10^7$. If we average the
parameter $T$ over the depth of the convective zone, we obtain $T
\approx 5$, see \cite{ketal03}.

\begin{figure*}
\begin{center}
\includegraphics[width=0.48\textwidth]{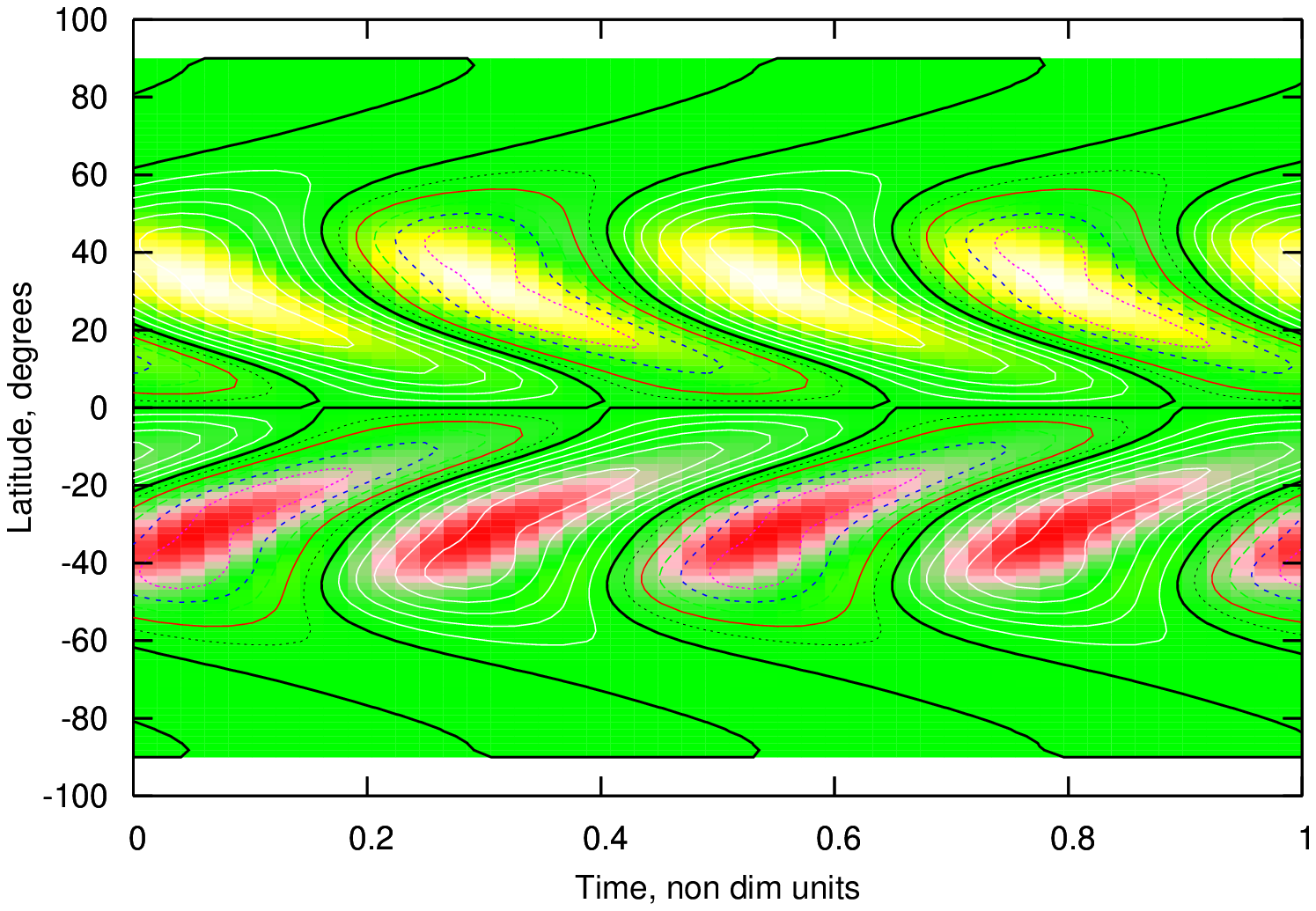}
\includegraphics[width=0.48\textwidth]{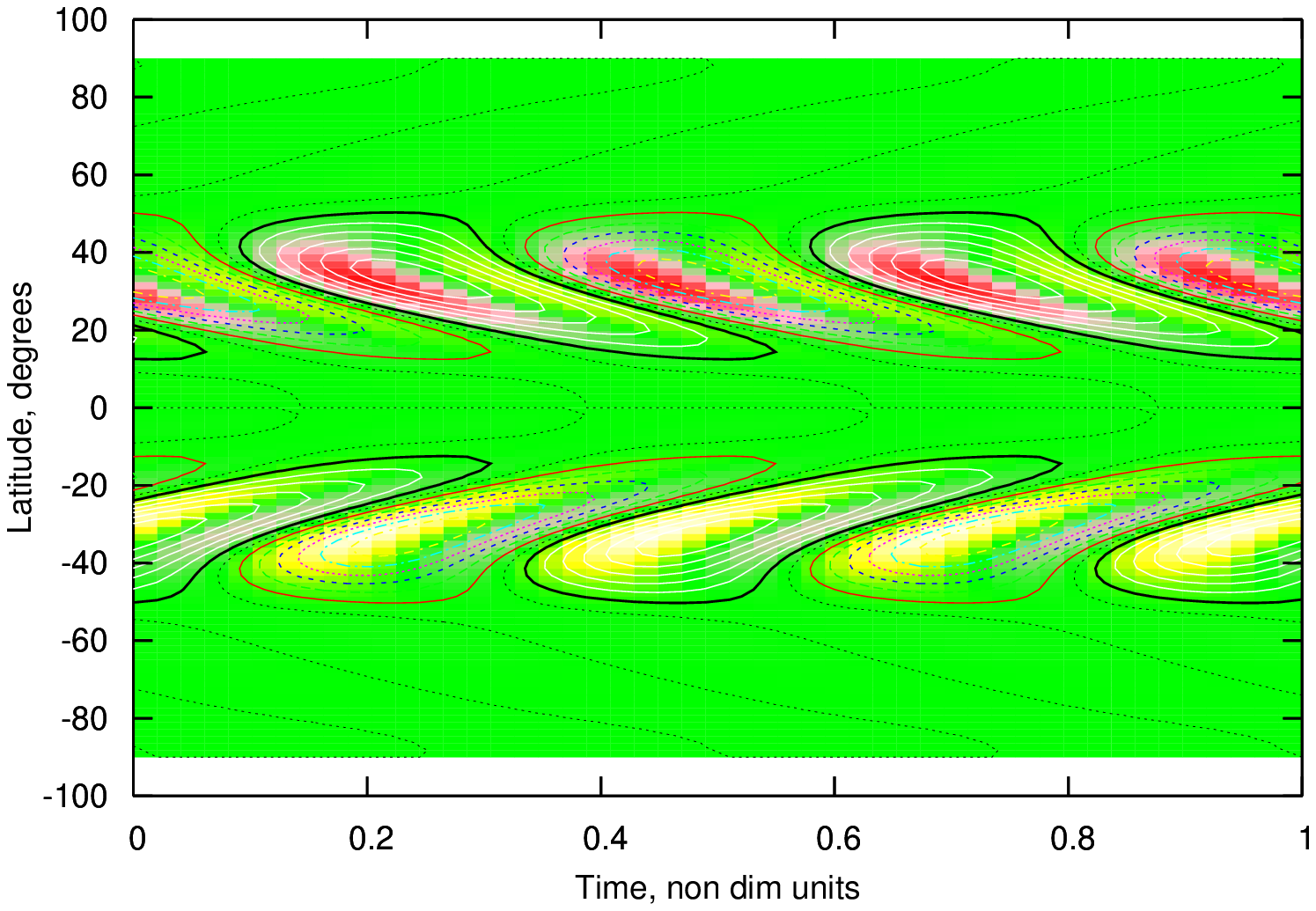}
\caption{Current helicity of active regions
estimated as $-{\bf A}\cdot{\bf B}$ [see
Eq.~(\ref{RRR5})], overlaid on toroidal field for
the primitive dynamo model: left panel - deep
layer, right - surface layer. $C_\alpha=-6.5$,
$C_\omega=6\cdot10^4$, $R_v=0$ (i.e. no
meridional circulation), no buoyancy. The
diffusivity constant  $\eta_{\rm r}=0.5$ and the
bottom of the computational region is at
$r_0=0.64$. The colour palette is hereafter
chosen as follows: yellow is positive, red is
negative, and  green is zero.} \label{Fig1}
\end{center}
\end{figure*}

We start with the results for the primitive model.
Fig.~\ref{Fig1} presents the current helicity butterfly diagrams
overlaid on those for the toroidal field. We estimate this
quantity based on the idea that
the observed current helicity in active
regions is expected to trace $-{\bf A} \cdot {\bf B}$,
and so we plot in this section $-{\bf A
\cdot B}$.

\begin{figure*}
\begin{center}
\includegraphics[width=0.48\textwidth]{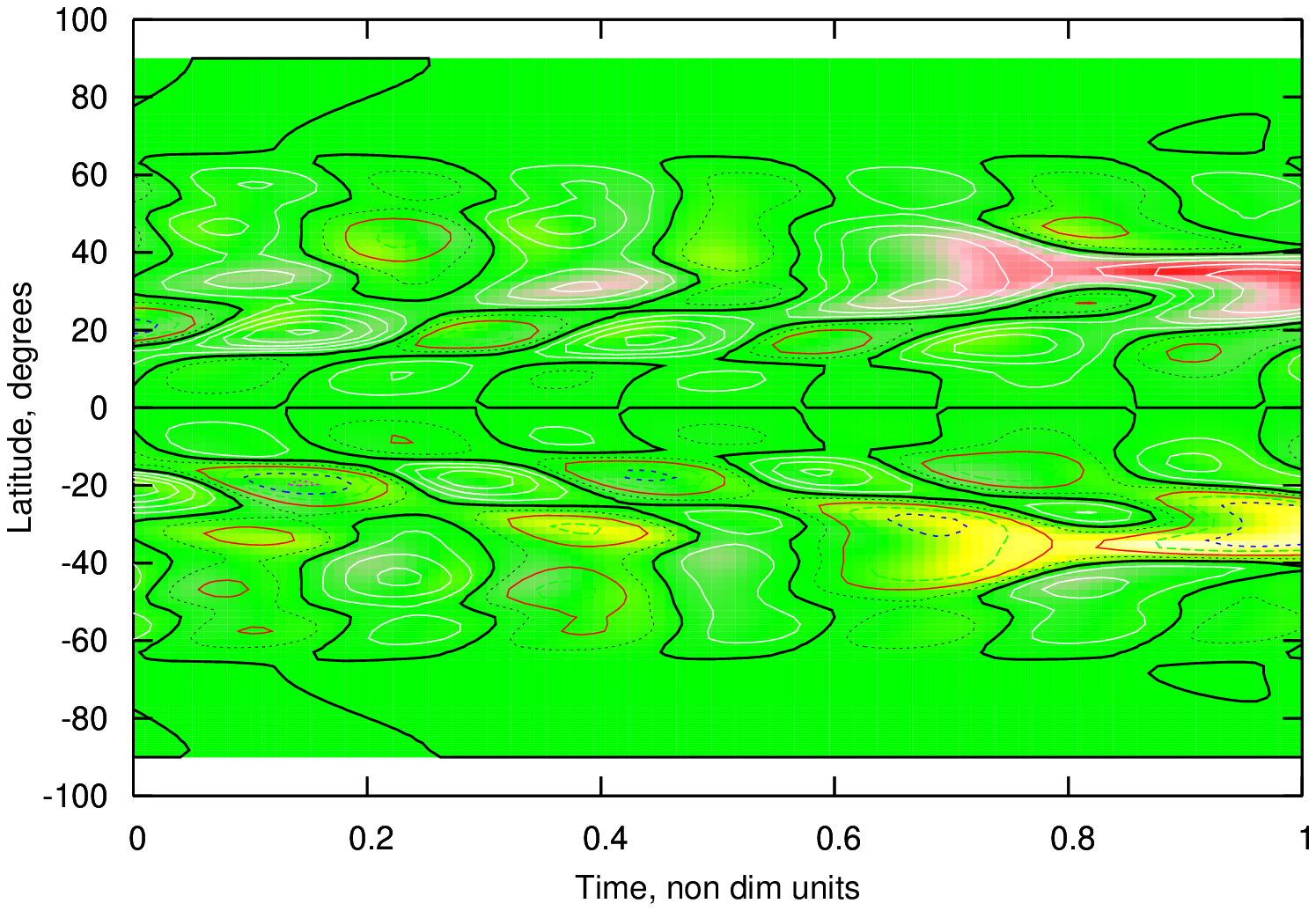}
\includegraphics[width=0.48\textwidth]{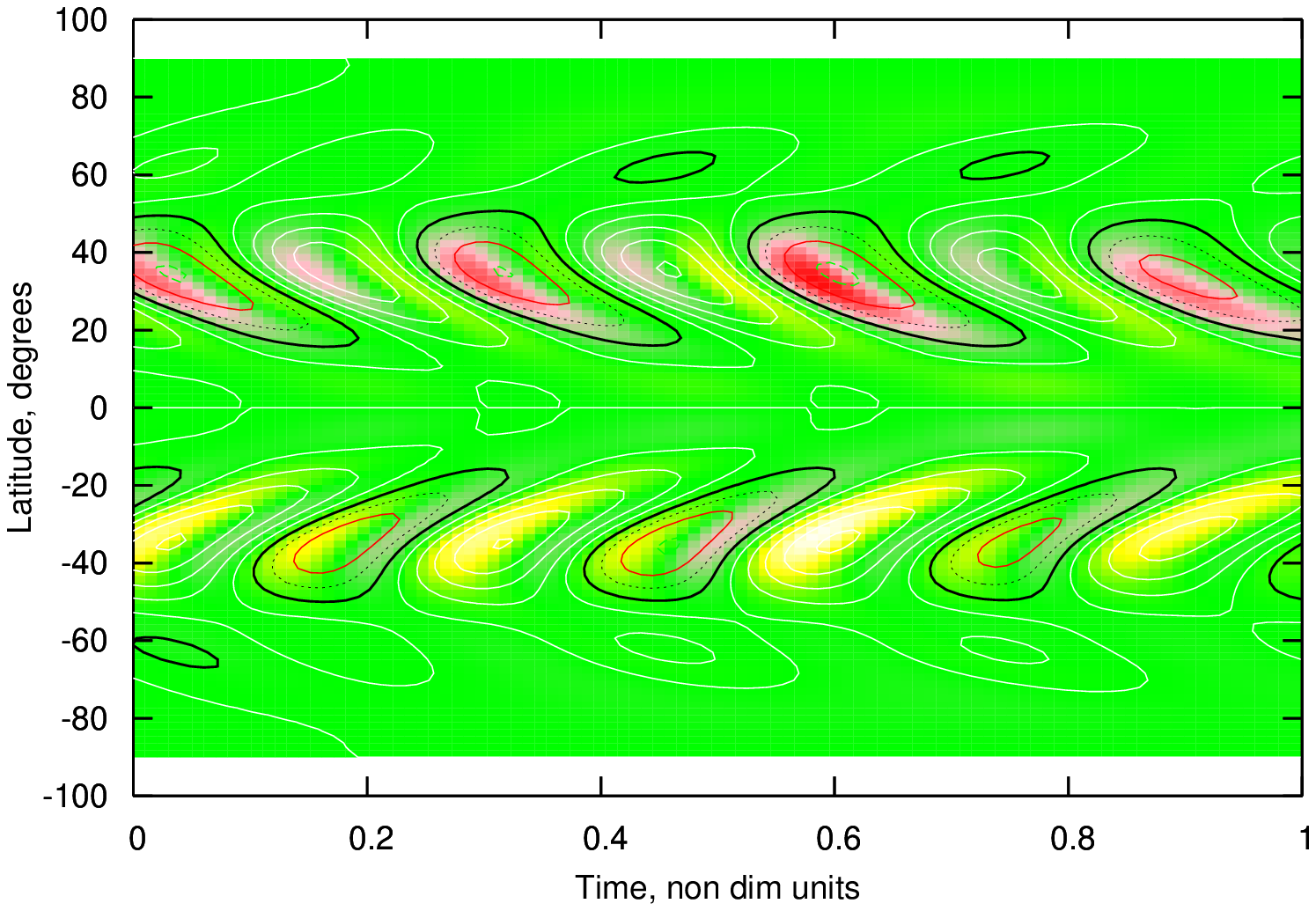}
\caption{Current helicity of active regions
estimated as $-{\bf A}\cdot{\bf B}$ [see
Eq.~(\ref{RRR5})], overlaid on toroidal field for
the primitive dynamo model with enhanced dynamo
activity in the overshoot layer: left panel -
deep layer, right - surface layer. The values of
$C_\alpha$, $C_\omega$, $R_v$ and $r_0$ are the
same as in the previous figure but diffusivity
contrast $\eta_{\rm r}=0.1$. } \label{Fig2}
\end{center}
\end{figure*}

We see that the plots successfully represent main feature of the
observed helicity patterns.
The pattern presented in Fig.~\ref{Fig1} is quite typical for the
model. Of course, one can choose a set of dynamo governing
parameters which is less similar to the observations. For example
one can concentrate magnetic fields in the deep layer of convective
zone (say, in the overshoot layer) by choosing a reduction
$\eta_{\rm r} = 0.1$ in the nominal "overshoot layer", instead of
$\eta_{\rm r} = 0.5$ (our standard case), as used in
Fig.~\ref{Fig1}. This tends to make the helicity wave in the
overshoot layer look more like a standing wave, but however keeps
the main features of surface diagram (Fig.~\ref{Fig2}). The highly
anharmonic standing patterns of the butterfly diagram that were discussed
as a possible option for some stars, see \cite{betal06}, look however
irrelevant for the solar case.

Of course, the helicity pattern in the butterfly
diagram obtained in the models for particular
choices of dynamo governing parameters can be
slightly different from the observed helicity
patterns. \cite{xetal09} demonstrated that
meridional circulation can be used to make the
simulated pattern resemble more closely that
observed.

\begin{figure*}
\begin{center}
\includegraphics[width=0.48\textwidth]{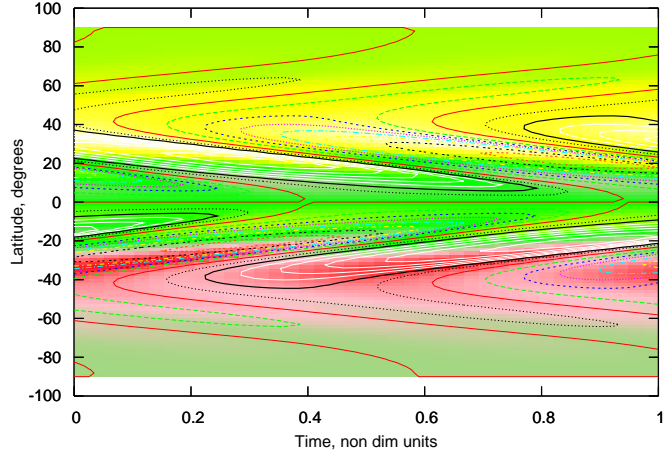}
\caption{Small-scale current helicity $\chi_c$
overlaid on toroidal field contours for the
dynamo model based on helicity balance, near the
middle radius of the dynamo region.
$C_\alpha=-5$, $C_\omega=6\cdot10^4$, $R_v=10$
(i.e. with meridional circulation), buoyancy
parameter $\gamma=1$. The plots are for
fractional radius $0.84$.} \label{Fig3}
\end{center}
\end{figure*}

\begin{figure*}
\begin{center}
\includegraphics[width=0.48\textwidth]
{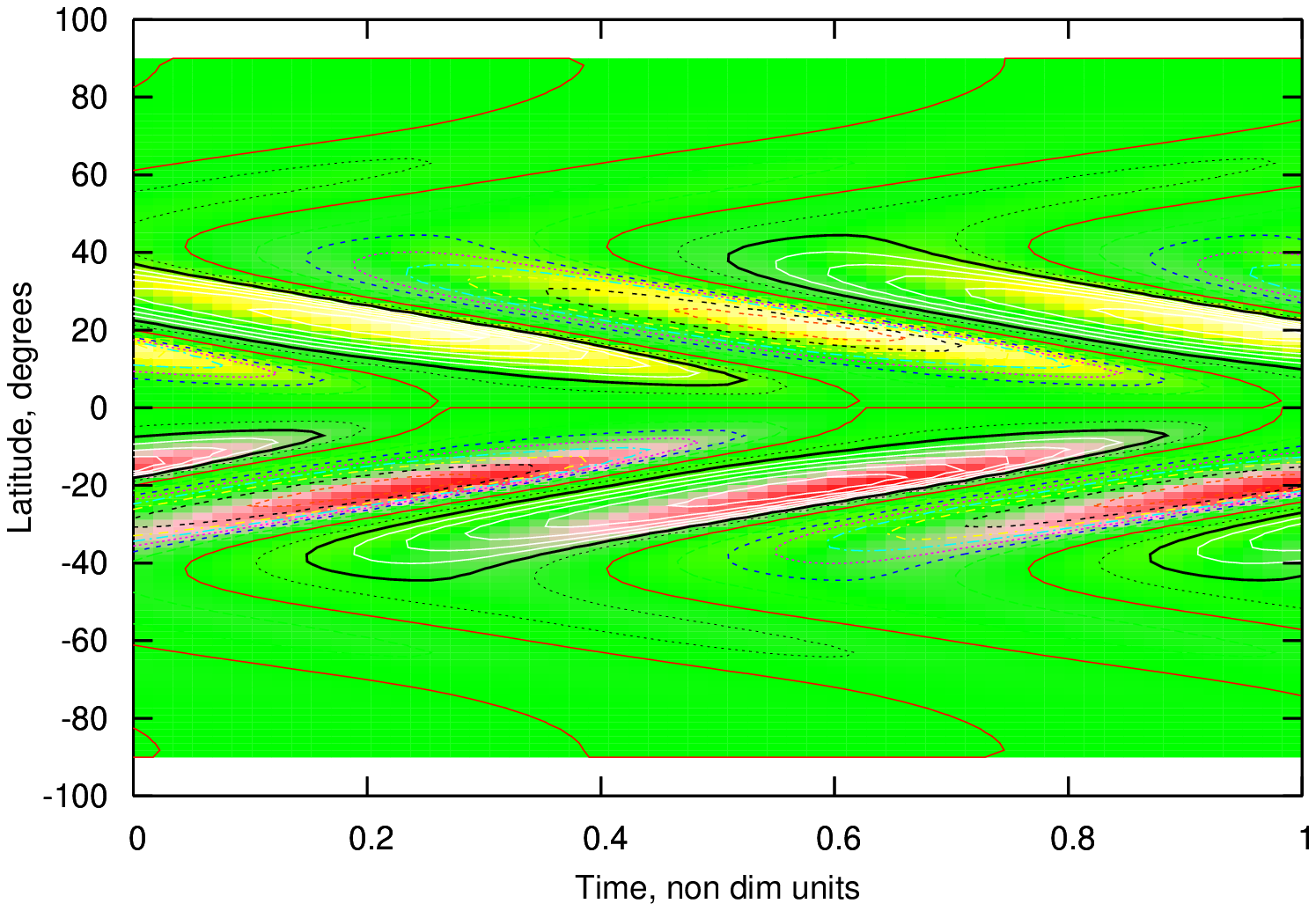}
\includegraphics[width=0.48\textwidth]
{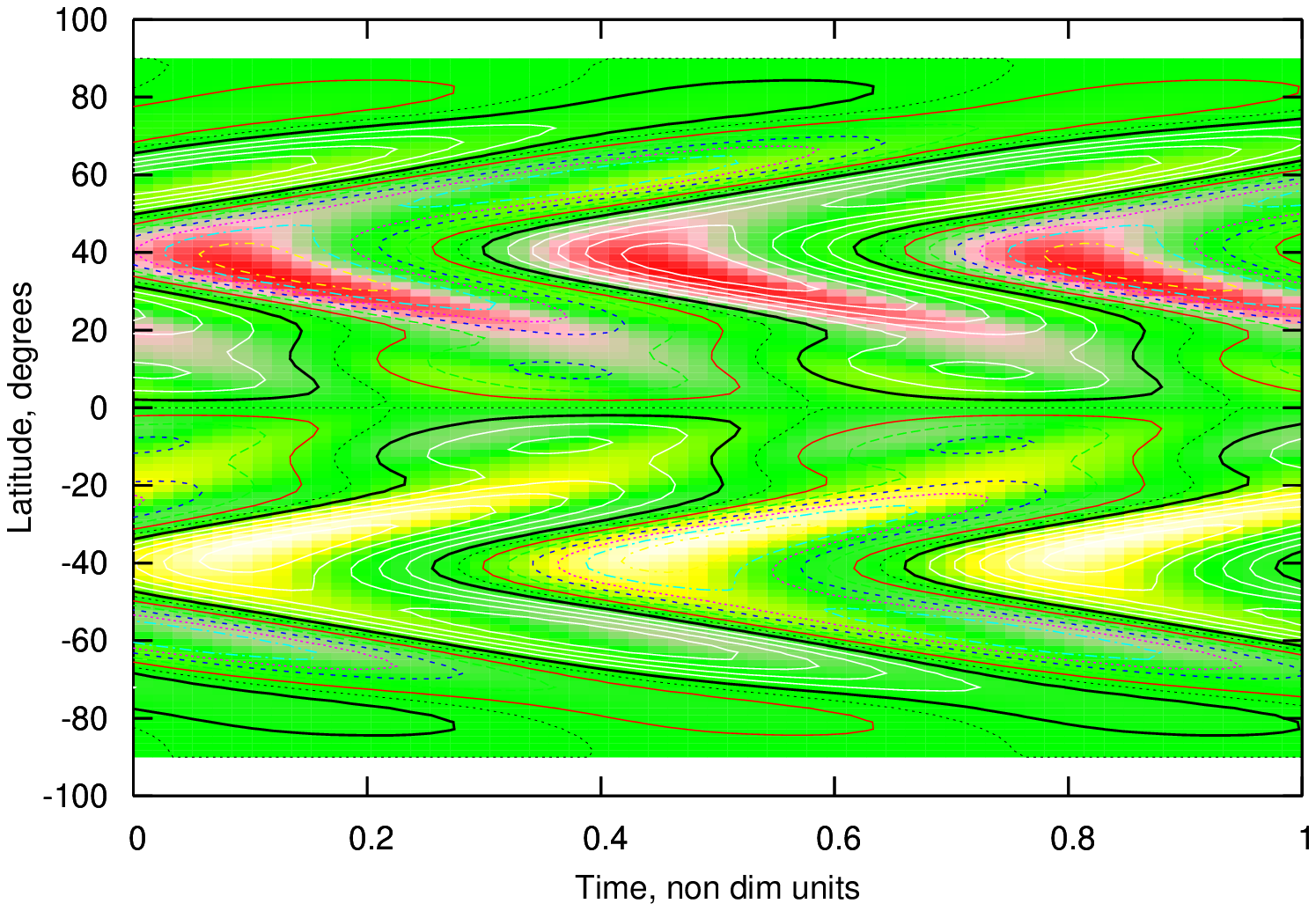} \caption{Current
helicity of active regions estimated  as $-{\bf
A}\cdot{\bf B}$ [see Eq.~(\ref{RRR5})], overlaid
on toroidal field contours for the dynamo model
based on helicity balance, near the middle radius
of the dynamo region ($r=0.84$, left panel) and
near the surface ($r=0.96$, right panel) for
$C_\alpha=-5$, $C_\omega=6\cdot10^4$, $R_v=10$
(i.e. with meridional circulation), and buoyancy
parameter $\gamma=1$. } \label{Fig4}
\end{center}
\end{figure*}

We produced the same type of plots for models
based on helicity conservation Fig.~\ref{Fig4}.
We also present in Fig.~\ref{Fig3} the
small-scale current helicity $\chi_c$. Here the
migration diagrams are presented for the middle
radius of the dynamo region: nearer the surface
$\chi_c$ displays only relatively weak
vacillatory behaviour. We see that the
small-scale current helicity is strongly
concentrated in middle latitudes and helicity
oscillations which are present in the model are
almost invisible on the background of the
intensive belt of constant helicity in middle
latitudes. We doubt that such oscillations would
be observable. We stress that, if this model
produces any travelling helicity pattern, it is
situated in the deep layers only. The pattern
usually is much more similar to that presented in
the right hand panel of Fig.~\ref{Fig2} rather
than to a travelling wave such as presented in
Fig.~\ref{Fig1}.

Taking our models as a whole, determination of
further details of the  location of the dynamo
layer or the helicity sign inversion  at some
latitudes and phases of the solar cycle is beyond
the scope of this investigation. Such a study
would require a more complex simulation, which we
hope to perform in the future.

The numerical models used here are extremely
simple and cannot be expected to sample the full
range of solutions that are accessible to a
three-dimensional, highly stratified, extremely
turbulent system with complex and unknown
boundary conditions. In the present paper we have
only shown the theoretical possibility that the
current helicity observed in solar active regions
may trace the magnetic helicity of the
large-scale dynamo generated field. We
demonstrated that the results obtained with our
simplified dynamo model are compatible with the
observations. At the same time we note that such
computationally complex and expensive simulations
are still far from being able to reproduce the
range of observed solar phenomena, and that for
the immediate future mean field models will
continue to play an important role. Indeed, they
may provide more immediate physical insight.

The simple models which we have considered here
were not intended to reproduce very fine details
of the spatial-time distribution of helicity
observable in the form of butterfly diagrams. We
may again note that our modelling of helicity in
the solar convective zone and active regions is
still too simplified to  be able to detect more
detailed properties of the helicity dynamics.

While we feel that it is remarkable that there is
a class of kinematic models that are able to
reproduce the current helicity observations from
vector magnetograms, we may again note that some
contribution to the observed current helicity may
be due to the surface effects caused by the
differential rotation \citep{MB05,YM09} and not
from the dynamo.

\section{Conclusions}

We conclude that current helicity of magnetic field in active regions
is a tracer of the magnetic helicity of  the large-scale magnetic field
in the solar interior. We believe that this provides a unique option
for tracing this quantity, which is very important for the solar dynamo.
According to the observational data \citep{zetal10}, the  current
helicity in active regions is mainly negative in the Northern
hemisphere. Numerical models give a negative value for
$-\bf A \cdot B$ in the surface layer of the convective zone
in the  Northern hemisphere.

Summarizing, we conclude that the current
helicity of the magnetic field in active regions
is expected to have the opposite sign to ${\bec{
A}} {\bf \cdot} {\bec{ B}}$, evaluated at the
depth at which the active region originates.
Thus, the models presented here are consistent
with the interpretation that the mechanism
responsible for the sign of the observed helicity
operates near the solar surface, cf. e.g.,
\cite{k10}.  The mechanism of formation of the
current helicity in active regions still requires
further investigation.

\acknowledgments

We acknowledge the NORDITA dynamo programs of
2009 and 2011 for providing a stimulating
scientific atmosphere.  This collaborative work
has been benefited from discussions at the
Helicity Thinkshop meeting in Beijing in 2009.
The work has been supported by Chinese grants
NSFC 10733020, 10921303, 41174153, 11103037,
NBRPC 2006CB806301, CAS  KLCX2-EW-T07 and joint
NNSF-RBFR grant 10911120051. K.K., D.S., N.K. and
D.M. would like to acknowledge support from the
National Astronomical Observatories of China
during their visits, K.K. and D.S. are thankful
to the joint project of NNSF of China and RFBR of
Russia no. 08-02-92211, to RFBR grant
10-02-00960, to Visiting Professorship Programme
2009J2-12, as well as joint Chinese-Russian
collaborative project from CAS.

\appendix

\section{Current helicity versus the magnetic helicity}
\label{CH-MH}

Here we relate the mean current helicity $\langle
{\bf B}^{\rm ar} {\bf \cdot} \bec{\rm curl}
\,{\bf B}^{\rm ar} \rangle$  with the magnetic
helicity $\langle {\bf A}^{\rm ar}{\bf \cdot}{\bf
B}^{\rm ar}\rangle$. First we rewrite the mean
current helicity from first principles with the
use of permutation tensors as
\begin{eqnarray}
\langle {\bf B}^{\rm ar} {\bf \cdot} \bec{\rm curl}
\,{\bf B}^{\rm ar} \rangle = \varepsilon_{mpq} \,
\varepsilon_{mij} \lim_{{\bf x} \to {\bf y}}
\, {\nabla}_p^{\bf x} \, {\nabla}_i^{\bf y} \langle
A^{\rm ar}_q({\bf x}) B^{\rm ar}_j({\bf y}) \rangle
= \lim_{{\bf x} \to {\bf y}} \, \big[(\bec{\nabla}^{\bf x}
\cdot \bec{\nabla}^{\bf y})
\, \langle {\bf A}^{\rm ar}({\bf x}) \cdot {\bf B}^{\rm ar}({\bf y})
\rangle - {\nabla}_p^{\bf x} \, {\nabla}_q^{\bf y}
\langle A^{\rm ar}_q({\bf x}) B^{\rm ar}_p
({\bf y}) \rangle \big] ,
\nonumber\\
\label{AP1}
\end{eqnarray}
where $\varepsilon_{ijn}$ is the fully
antisymmetric Levi-Civita tensor, $R_\odot$ is
the solar radius, $L_{\rm ar}$ is the spatial
scale  of an active region and ${\bf r} = {\bf x}
- {\bf y}$. The use of the full tensor notation
and limits are needed here in order to separate
the large-scale and small-scale variables and to
obtain a simple final answer in scalar form when
there is separation  of scales. Since ${\bf r} =
{\bf x} - {\bf y}$ is a small-scale variable,
${\bf R} = ({\bf x} + {\bf y})/2$ is a
large-scale variable, the derivatives
\begin{eqnarray*}
{\nabla}_p^{\bf x} \equiv {\partial \over \partial x_p}
= {\partial \over \partial r_p} + {1\over 2} {\partial
\over \partial R_p} = - {\nabla}_p^{\bf y} + {\partial
\over \partial R_p},
\quad \quad \quad
{\nabla}_p^{\bf y} \equiv {\partial \over \partial y_p}
= - {\partial \over \partial r_p} + {1\over 2} {\partial
\over \partial R_p} = - {\nabla}_p^{\bf x} + {\partial \over \partial R_p} .
\end{eqnarray*}
This implies that
\begin{eqnarray*}
\bec{\nabla}^{\bf x} \cdot \bec{\nabla}^{\bf y} = - \left({\partial^2
\over \partial {\bf r}^2} - {1\over 4} {\partial^2 \over \partial {\bf R}^2}\right),
\quad \quad \quad
{\nabla}_p^{\bf x} {\nabla}_q^{\bf y} = {\nabla}_p^{\bf y} {\nabla}_q^{\bf x}
- {\nabla}_p^{\bf y} {\partial \over \partial R_q} - {\nabla}_q^{\bf x}
{\partial \over \partial R_p} + {\partial^2 \over \partial R_p \partial R_q} .
\end{eqnarray*}
We take into account that div$\, {\bf B}^{\rm ar}
=0$ [i.e. ${\nabla}_p^{\bf y} B^{\rm ar}_p({\bf
y})=0$]  and div$\, {\bf A}^{\rm ar} =0$ [i.e.
${\nabla}_q^{\bf x} A^{\rm ar}_q({\bf x})=0$]. We
also take into account that the characteristic
scale of an active region is small compared with
the thickness of the convection zone or the
radius of the Sun. This implies that
\begin{eqnarray*}
{\nabla}_p^{\bf x} \, {\nabla}_q^{\bf y} \langle A^{\rm ar}_q({\bf x})
 B^{\rm ar}_p({\bf y}) \rangle = {\partial^2 \over \partial R_p \partial R_q}
 \langle A^{\rm ar}_q({\bf x}) B^{\rm ar}_p({\bf y}) \rangle \sim
  O\biggl({L^2_{\rm ar} \over R_\odot^2}\biggr),
\end{eqnarray*}
and therefore this term vanishes. This yields
\begin{eqnarray}
\langle {\bf B}^{\rm ar} {\bf \cdot} \bec{\rm curl} \,{\bf B}^{\rm ar}
\rangle = - \biggl({\partial^2 \over \partial r_p \partial r_p} \langle
{\bf A}^{\rm ar}{\bf \cdot}{\bf B}^{\rm ar} \rangle \biggr)_{{\bf r}
\to 0}+ O\biggl({L^2_{\rm ar} \over R_\odot^2}\biggr),
\label{AP2}
\end{eqnarray}
Now we take into account that the second derivative
of the correlation function
\begin{eqnarray*}
\biggl({\partial^2 \over \partial r_p \partial r_p} \langle
{\bf A}^{\rm ar}{\bf \cdot}{\bf B}^{\rm ar} \rangle
\biggr)_{{\bf r} \to 0}
\end{eqnarray*}
should be negative since as ${\bf r} \to 0$ the
correlation function has a maximum.  Thus, we
finally obtain
\begin{eqnarray}
\langle {\bf B}^{\rm ar} {\bf \cdot} \bec{\rm curl} \,{\bf B}^{\rm ar} \rangle
\approx {1\over L^2_{\rm ar}} \, \langle {\bf A}^{\rm ar}{\bf \cdot}{\bf B}^{\rm ar} \rangle
+ O\biggl({L^2_{\rm ar} \over R_\odot^2}\biggr) .
\label{AP3}
\end{eqnarray}
Similar calculations relating the current
helicity and the magnetic helicity in ${\bf
k}$-space can be found in Appendix C of
\cite{KR99}.

\section{Detailed description of the dynamo model}
\label{Dyn-mod}

The flux of the small-scale magnetic helicity is chosen in the form
\begin{eqnarray}
\bec{\cal F} &=& \eta_{_{A}}(B) \, B^2 \, \{C_1 \, \bec{\nabla}
[\chi^v \, \phi_v(B)] + C_2 \, \chi^v \, \phi_v(B) \,
\bec{\Lambda}_{\rho} \}
- C_3 \, \kappa \, \bec{\nabla} \, \chi^c  \;,
\label{flux}
\end{eqnarray}
with $\bec{\Lambda}_\rho = - \bec{\nabla} \rho /
\rho$ and $\kappa$ is the coefficient of
turbulent diffusion of small-scale magnetic
helicity (see below).

The quenching functions $\Phi_{v}(B)$ and $\Phi_{m}(B)$ appearing
in the nonlinear $\alpha$ effect are given by
\begin{eqnarray}
\Phi_{v}(B) &=& {1\over 7} [4 \phi_{m} (B) + 3 L(\sqrt{8} B)] \;,
\label{L5} \\
\Phi_{m} (B) &=& {3 \over 8 B^2} \biggl[1 - {\arctan (\sqrt{8} B)
\over \sqrt{8} B} \biggr] \;
\label{L6}
\end{eqnarray}
\citep{RK00}, where $ L(y) = 1 - 2 y^{2} +
2 y^{4} \ln (1 + y^{-2})$.

The nonlinear turbulent magnetic diffusion coefficients for the
mean poloidal and toroidal magnetic fields, $\eta_{_{A}}(B)$ and
$\eta_{_{B}}(B)$, are given in dimensionless form by
\begin{eqnarray}
\eta_{_{A}}(B) &=& A_{1}(4 B) + A_{2}(4 B) \;,
\label{D1} \\
\eta_{_{B}}(B) &=& A_{1}(4 B) + {3 \over 2} [2 A_{2}(4 B) -
A_{3}(4 B)] \;,
\label{D2}
\end{eqnarray}
\citep{RK04}.
For the case of weak magnetic field the turbulent
diffusion coefficients are (in units of the reference value
$\eta_{_{T0}}$)

\begin{equation}
\eta_{_{A}} = 1 - {{96} \over 5} \, B^2 \;, \quad \eta_{_{B}} = 1
- 32 B^2, \label{smalB}
\end{equation}
while for strong magnetic fields the scaling is

\begin{equation}
\eta_{_{A}} = {1 \over {8 B^2}}, \quad \quad \eta_{_{B}} = {1
\over {3 \sqrt{2} B}} \; . \label{largeB}
\end{equation}
The transition from one asymptotic form to the other can be
thought of as occurring in the vicinity of $B\sim B_{\rm eq}/4$.

The nonlinear drift velocities of poloidal
and toroidal mean magnetic fields, $\bec{V}^{A}(B)$ and
$\bec{V}^{B}(B)$, are given in dimensionless form by

\begin{eqnarray}
\bec{V}^{A}(B) &=& V_1(B) {\bec{\Lambda}_B \over 2} + {V_2(B)
\over r} \, (\bec{e}_{r} + \cot \theta \, \bec{e}_{\theta}) +
\bec{V}_\rho(B)\;,
\nonumber\\
\label{D3}\\
\bec{V}^{B}(B) &=& {V_3(B) \over r} \, (\bec{e}_{r} + \cot \theta
\, \bec{e}_{\theta}) + \bec{V}_\rho(B) \;, \label{D4}
\end{eqnarray}
where
\begin{eqnarray*}
V_1(B) &=& {3 \over 2} A_{3}(4 B) - 2 A_{2}(4 B) \;,
\\
V_2(B) &=& {1 \over 2} A_{2}(4 B) \;,
\\
V_3(B) &=& {3 \over 2} [ A_{2}(4 B) - A_{3}(4 B)] \;,
\\
\bec{V}_\rho(B) &=& {1 \over 2} \bec{\Lambda}_\rho [-5A_{2}(4 B) +
{3 \over 2} A_{3}(4 B)].
\end{eqnarray*}
The asymptotic formulae for these velocities are given by
\begin{eqnarray*}
\bec{V}^{A} &=& {32 \over 5} \, B^{2} \, \biggl[ \,
\bec{\Lambda}_B + 3 \bec{\Lambda}_\rho - {{\bec{e}_{r} + \cot
\theta \, \bec{e}_{\theta} }\over {r} }\biggr] \;,
\\
\bec{V}^{B} &=& {32 \over 5} B^2 \biggl[ 3\bec{\Lambda}_\rho-
{\bec{e}_{r} + \cot \theta \, \bec{e}_{\theta} \over {r}}\biggr]
\;
\end{eqnarray*}
for a weak magnetic field, and
\begin{eqnarray*}
\bec{V}^{A} &=& - {1 \over 3 \sqrt 8 B} \biggl[\bec{\Lambda}_B + 2
{\bec{e}_{r} + \cot \theta \, \bec{e}_{\theta} \over {r}} \biggr]
+ {5 \over 16 B^2} \bec{\Lambda}_\rho \;,
\\
\bec{V}^{B} &=& {4 \over 3 \sqrt 8 B} {\bec{e}_{r} + \cot \theta
\, \bec{e}_{\theta} \over {r}} + {5 \over 16 B^2}
\bec{\Lambda}_\rho \;
\end{eqnarray*}
for strong fields. Here $\bec{\Lambda}_B = (\bec{\nabla}
\bec{B}^2) / \bec{B}^2 $, $\bec{e}_r$ and $\bec{e}_\theta$ are
unit vectors in the $r$ and $\theta$ directions of spherical polar
coordinates, $[\bec{\Lambda}_\rho]_r=-d\ln \rho/dr$, and
$[\bec{\Lambda}_B]_r=d\ln B^2/dr$. See, for details,
Rogachevskii \& Kleeorin [2004, Eqs.~(18), (19), (22)-(24)],
which have been rewritten here in spherical geometry.

The functions $ A_{k}(y) $ are
\begin{eqnarray*}
A_{1}(y) &=& {6 \over 5} \biggl[{\arctan y \over y} \, \biggl(1 +
{5 \over 7 y^{2}} \biggr) + {1 \over 14} L(y) - {5 \over 7y^{2}}
\biggr] \;,
\\
A_{2}(y) &=& - {6 \over 5} \biggl[{\arctan y \over y} \, \biggl(1
+ {15 \over 7 y^{2}} \biggr) - {2 \over 7} L(y) - {15 \over
7y^{2}} \biggr] \;,
\\
A_{3}(y) &=& - {2 \over y^{2}} \biggl[ {\arctan y \over y} \,
(y^{2} + 3) - 3 \biggr] \; .
\end{eqnarray*}

The nonlinear quenching of the turbulent magnetic diffusion of the
magnetic helicity is given by
\begin{eqnarray}
\kappa(B) = {1 \over 2} \biggl[1 + A_{1}(4B) + {1 \over 2}
A_{2}(4B) \biggr] \; . \label{D10}
\end{eqnarray}
The coefficient of turbulent diffusion of magnetic
helicity $\kappa$ also has a dependence on $B$, namely $\kappa(B)
= 1 - 24 B^2/5$ for weak magnetic field and

\begin{equation}
\kappa(B) = {1 \over 2} \biggl( 1 + {{3 \pi} \over {40 B}} \biggr)
\end{equation}
in the strong field limit.

\end{document}